\documentclass[doublecol]{epl2}
\usepackage{amsmath}
\usepackage{amsfonts}
\usepackage{amssymb}
\usepackage{hyperref}
\title{Fluctuation induced non-canonical BCS states: A mechanism for pseudogap}
\shorttitle{Non-canonical BCS states}
\author{Lei Gu \thanks{E-mail: \email{leigu@ymail.com}}}
\shortauthor{Lei Gu} \institute{Research Center for Quantum
Manipulation and Department of Physics, Fudan University, Shanghai
200433, People's Republic of China}

\pacs{05.30.-d}{Quantum statistical mechanics}
\pacs{74.20.Fg}{BCS
theory and its development}
\pacs{03.65.Yz}{Decoherence; open
systems; quantum statistical methods}

\abstract{We pose the question of what effect the statistical
fluctuation causes if it induces a non-unitary evolution. We apply
this idea to the BCS model and study fluctuation around the
mean-field average. We find that, dynamics of the thermalization
influences the equilibrium besides the non-unitary evolution, and
the resulting equilibrium state is no longer the canonical one. The
pseudogap phenomenon can exist in this model.}

\begin{document}

\maketitle

\section{Introduction}
Consider equilibrium states of a system governed by master equation
\begin{equation}
\frac{\textrm{d}\rho}{\textrm{d}t}=-i[\hat{H},\rho]-\eta[\hat{A},[\hat{A},\rho]],\label{master}
\end{equation}
where $\hat{A}$ is an operator of the system. The equation is a
simplified Lindblad equation~\cite{lindb} with the operators
specified as a Hermitian one. If an energy eigenstate $|
E\rangle\langle E|$ is not an eigenstate of $\hat{A}$ , it is in
general not time-independent, $\upd(| E\rangle\langle E|)/\upd
t\neq0$. In this case, the canonical state $\rho\propto
\textrm{e}^{-\hat{H}/k_{B}T}$ can not be an equilibrium state, and a
non-canonical one is required. Suppose the goal of thermal
relaxation is the canonical equilibrium ensemble\footnote{The
argument is given in the Appendix.}. The thermalization ceases when
the goal is achieved. When the system is in a non-canonical
equilibrium state, however, the thermalization still works. So the
dynamics of thermalization is also a factor in determination of the
equilibrium.

The BCS theory is a non-interaction quasiparticle model. The
subadditivity~\cite{pere} of entropy for a composite system
$\rho^{AB}$ indicates $S(\rho^{AB})\leqslant
S(\textrm{Tr}_{A}[\rho^{AB}]\otimes \textrm{Tr}_{B}[\rho^{AB}])$.
When there is no interaction between the two subsystems, this
inequality can be generalized to the free energy as
$F(\rho^{AB})\geqslant F(\textrm{Tr}_{A}[\rho^{AB}]\otimes
\textrm{Tr}_{B}[\rho^{AB}])$. As long as $\hat{A}$ does not mix
states of different momentum, according to the principle of minimum
free energy, a BCS state after thermal relaxation must take the form
$\rho=\otimes_{\bm{k}}\rho_{\bm{k}}$~\cite{leigu}, where
$\rho_{\bm{k}}$ is a density matrix in the subspace of momentum
$\bm{k}$. So we can handle each subspace separately, which allow us
to study the energy variation caused by a gap for each momentum. We
assume that a gap emerges only in the subspaces where the energy is
reduced by the gapping. For the gapped momentums, we adopt the
Cooper approximation, a constant $\Delta_{\bm{k}}$.

Here, a subspace of momentum $\bm{k}$ means the one spanned by the
ground state component, two excitons and the double exciton:
\begin{eqnarray}
\nonumber&&|0\rangle=|\Phi_{\bm{k}}\rangle=u_{\bm{k}}|\
\rangle+v_{\bm{k}}c_{\bm{k}\uparrow}^{+}c_{-\bm{k}\downarrow}^{+}|\
\rangle,\\
\nonumber&&|1\rangle=b_{\bm{-k}\downarrow}^{+}|\Phi_{\bm{k}}\rangle=c_{-\bm{k}\downarrow}^{+}|\
\rangle,\
|2\rangle=b_{\bm{k}\uparrow}^{+}|\Phi_{\bm{k}}\rangle=c_{\bm{k}\uparrow}^{+}|\
\rangle,\\
&&|3\rangle=b_{\bm{k}\uparrow}^{+}b_{-\bm{k}\downarrow}^{+}|\Phi_{\bm{k}}\rangle=-v_{\bm{k}}^{*}|\
\rangle+u_{\bm{k}}^{*}c_{\bm{k}\uparrow}^{+}c_{-\bm{k}\downarrow}^{+}|\
\rangle,
\end{eqnarray}
where $|\ \rangle$ is the bare vacuum, $b_{\bm{k}\uparrow}^{+},
b_{-\bm{k}\downarrow}^{+}$ the Bogoliubov creation operators and
$c_{-\bm{k}\downarrow}^{+},c_{\bm{k}\uparrow}^{+}$ creation
operators of an electron. The subspace spanned by states of
$\bm{k}\downarrow$ and $-\bm{k}\uparrow$ are referred to by
$-\bm{k}$.

To have a sense of how we can deal with a subspace separately, let
us derive the Fermi-Dirac distribution for the BCS model according
to the principle of minimum free energy. A density operator in a
subspace can be expressed as
$\rho_{\bm{k}}=\sum_{mn}a_{mn}|m\rangle\langle n|$ with
$a_{mn}^{*}=a_{nm}$. Since the time-independent states of von
Neumman equation are energy eigenstates, an equilibrium state has
form $\rho_{\bm{k}}=\sum_{n=0}^{3}a_{nn}|n\rangle\langle n|$. We
assign variables as $a_{00}=p_{-},
a_{33}=p_{+},a_{11}=a_{22}=(1-p_{-}-p_{+})/2$ and set the energy of
the ground states zero. Then we have
$E^{1}=E^{2}=E^{3}/2=E_{\bm{k}}=\sqrt{\omega^{2}+\Delta^{2}},\omega=\epsilon_{\bm{k}}-\epsilon_{F}$.
By minimizing the free energy, we obtain the probabilities
\begin{equation}
p_{\mp}=(\textrm{e}^{\mp\beta E_{\bm{k}}}+1)^{-2},
\end{equation}
where $\beta=k_{B}T$. The Fermi-Dirac distribution is given by
$(1-p_{-}+p_{+})/2=(\textrm{e}^{\beta E_{\bm{k}}}+1)^{-1}$

\section{Non-unitary evolution and thermalization}
A many-body Hamiltonian can not lead to non-unitary evolution by
itself, because no environment is involved. For a part of this
many-particle system, however, the other parts actually play a roll
as an environment with fluctuations, which can cause
decoherence~\cite{Zurek}. Consider the subspace of momentum
$\bm{k}$, Hamiltonian for which is given by
\begin{equation}
H_{\bm{k}}=E_{\bm{k}}(b_{\bm{k}\uparrow}^{+}b_{\bm{k}\uparrow}+b_{\bm{-k}\downarrow}^{+}
b_{\bm{-k}\downarrow})+P,
\end{equation}
where $P$ is the fluctuation around the mean field average
\begin{equation}
P=c_{\bm{k}\uparrow}^{+}c_{-\bm{k}\downarrow}^{+}(-g)\sum_{\bm{k}'}(c_{-\bm{k}'\downarrow}c_{\bm{k}'\uparrow}-\langle
c_{-\bm{k}'\downarrow}c_{\bm{k}'\uparrow}\rangle)+c.c.
\end{equation}
From
$c_{-\bm{k}\downarrow}c_{\bm{k}\uparrow}=(\Delta_{\bm{k}}/E_{\bm{k}})(1-n_{-\bm{k}\downarrow}-n_{\bm{k}\uparrow})$,
$P$ is actually a fluctuation concerning the particle number in
gapped states, and it sets off when a gap emerges. To have an
explicit expression of the non-unitary dynamics cause by this
fluctuation, we assume the fluctuation can be approximated by a
Gaussian noise. The trivial equality $\langle
c_{-\bm{k}'\downarrow}c_{\bm{k}'\uparrow}\rangle-\langle
c_{-\bm{k}'\downarrow}c_{\bm{k}'\uparrow}\rangle=0$ indicates that
the fluctuation is a non-biased noise. We specify the phase factor
so that $\Delta_{\bm{k}}$ is a real number. Then the subspace is
coupled to the noise though operator
$\hat{A}=c_{\bm{k}\uparrow}^{+}c_{-\bm{k}\downarrow}^{+}+c_{-\bm{k}\downarrow}c_{\bm{k}\uparrow}$
with coupling $-g$. In the ``interaction picture'' the non-unitary
evolution has a Lindblad form~\cite{Matsu}
\begin{equation}
\frac{\upd\rho_{\bm{k}}^{I}}{\upd
t}=-g^{2}\tau_{c}[\hat{A},[\hat{A},\rho_{\bm{k}}^{I}]]\label{nonuni}
\end{equation}
where $\tau_{c}\equiv\int_{0}^{\infty}f(t)f(0)dt$ is the
autocorrelation for $t=0$, at which the gap is opened up. In the
Schr\"odinger picture, Eq.~(\ref{nonuni}) takes the form of
Eq.~(\ref{master}).

Since eigenstates of $\hat{A}$ are different from the energy
eigenstates, we need to construct a thermalization model. Effect of
the thermalization is reduction of the free energy. To achieve the
canonical ensemble, we may expect a phenomenological description of
the thermalization as the following: for off-diagonal elements
$a_{mn},m\neq n$, the time-dependence obeys
\begin{equation}
\frac{\upd a_{mn}}{\upd t}=-f_{mn}(\rho_{\bm{k}})a_{mn},
\end{equation}
while diagonal elements $a_{nn}$ vary with time as
\begin{equation}
\frac{\upd a_{nn}}{\upd
t}=-h_{nn}(\rho_{\bm{k}})\frac{\partial{F}}{\partial{a_{nn}}}.\label{minimu}
\end{equation}
Here, $f_{mn}(\rho_{\bm{k}}),h_{nn}(\rho_{\bm{k}})$ are positive
functions of $\rho_{\bm{k}}$ representing thermalization properties
of the system, and
\begin{equation}
F=a_{nn}E^{n}+Tk_{B}a_{nn}\ln{a_{nn}}
\end{equation}
is the free energy without considering the off diagonal elements.
For simplicity, we consider the functions two constants, $2f$ and
$2h$ respectively. The factor $2$ is included to simplify latter
notation.

The overall dynamics of our BCS model consists of three components,
that is, the von Neumman equation, the non-unitary evolution and the
thermal relaxation ($R$), i.e.
\begin{equation}
 \frac{\upd\rho_{\bm{k}}}{\upd
t}=-i[\hat{H},\rho_{\bm{k}}]-g^{2}\tau_{c}[\hat{A},[\hat{A},\rho_{\bm{k}}]]+R\label{overall}
\end{equation}
Without the non-unitary evolution, the off-diagonal elements vary as
\begin{equation}
a_{mn}(t)=a_{mn}(t_{0})e^{\textrm{i}(E^{n}-E^{m})t-ft}
\end{equation}
which implies that they approach zero through relaxation. Then the
von Neumman term gives zero because energy eigenstates are
commutable with the Hamiltonian. The master equation~(\ref{overall})
reduces to Eq.~(\ref{minimu}), where the time-independence is just
the condition for minimum free energy. The solution is the canonical
equilibrium state.

\section{The pseudogap}
The operator
$\hat{A}=c_{\bm{k}\uparrow}^{+}c_{-\bm{k}\downarrow}^{+}+c_{-\bm{k}\downarrow}c_{\bm{k}\uparrow}$
acts as:
\begin{eqnarray}\
\nonumber&&\hat{A}|\
\rangle=c_{\bm{k}\uparrow}^{+}c_{-\bm{k}\downarrow}^{+}|\ \rangle,\
\hat{A}c_{\bm{k}\uparrow}^{+}c_{-\bm{k}\downarrow}^{+}|\
\rangle=|\ \rangle,\\
&&\hat{A}c_{-\bm{k}\downarrow}^{+}|\
\rangle=\hat{A}c_{\bm{k}\uparrow}^{+}|\ \rangle=0.
\end{eqnarray}
Action of $[\hat{A},[\hat{A},\rho]]$ annihilate the state elements
except $|m\rangle\langle n|(m,n=0,3)$. The non-unitary evolution
impose no constraints on matrix elements $a_{mn}(m,n\neq0,3)$. Those
elements approach zero through thermal relaxation except $a_{11}$
and $a_{22}$. We only need to consider the following matrix
\begin{equation}
\rho_{\bm{k}}=\left(\begin{array}{cccc}
a_{00}&0&0&a_{03}\\
0&a_{11}&0&0\\
0&0&a_{22}&0\\
a_{30}&0&0&a_{33}\end{array}\right),
\end{equation}
and solve the equations given by coefficients of the six state
elements. Among the six equations, the two from $|1\rangle\langle
1|, |2\rangle\langle 2|$ have the same solution, and the two from
$|0\rangle\langle 3|, |3\rangle\langle 0|$ are equivalent because of
the Hermitian symmetry. So only four equations are independent.

It is more convenient to obtain the equilibrium state through
equations from the coefficients of $|\ \rangle\langle\
|,c_{\bm{k}\uparrow}^{+}c_{-\bm{k}\downarrow}^{+}|\ \rangle\langle\
|c_{-\bm{k}\downarrow}c_{\bm{k}\uparrow},|\ \rangle\langle\
|c_{-\bm{k}\downarrow}c_{\bm{k}\uparrow}$ and
$c_{-\bm{k}\downarrow}^{+}|\ \rangle\langle\
|c_{-\bm{k}\downarrow}$, which are given by
\begin{eqnarray}
\nonumber\textrm{i}u_{\bm{k}}v_{\bm{k}}(a_{00}-a_{33})-(h+2\eta)u_{\bm{k}}v_{\bm{k}}(a_{03}+a_{30})+\\
\eta(u_{\bm{k}}^{2}-v_{\bm{k}}^{2})(a_{00}-a_{33})+u_{\bm{k}}^{2}\frac{\partial{F}}{\partial{a_{00}}}+
v_{\bm{k}}^{2}\frac{\partial{F}}{\partial{a_{33}}}=0,\label{first}\\
\nonumber-\textrm{i}u_{\bm{k}}v_{\bm{k}}(a_{00}-a_{33})+(h+2\eta)u_{\bm{k}}v_{\bm{k}}(a_{03}+a_{30})-\\
\eta(u_{\bm{k}}^{2}-v_{\bm{k}}^{2})(a_{00}-a_{33})+v_{\bm{k}}^{2}\frac{\partial{F}}{\partial{a_{00}}}+
u_{\bm{k}}^{2}\frac{\partial{F}}{\partial{a_{33}}}=0,\label{secon}\\
\nonumber-\textrm{i}Eu_{\bm{k}}^{2}a_{03}-\textrm{i}Ev_{\bm{k}}^{2}a_{30}+\eta(a_{03}-a_{30})+\\
u_{\bm{k}}v_{\bm{k}}(\frac{\partial{F}}{\partial{a_{00}}}-
\frac{\partial{F}}{\partial{a_{33}}})=0,\\
\frac{\partial{F}}{\partial{a_{11}}}=0,\label{last}
\end{eqnarray}
where $\eta=g^{2}\tau_{c}$.

From summation of Eq.~(\ref{first}) and Eq.~(\ref{secon}), we have
\begin{equation}
\frac{\partial{F}}{\partial{a_{00}}}+
\frac{\partial{F}}{\partial{a_{33}}}=0.\label{free}
\end{equation}
Thus Eq.~(\ref{last}) holds according to
$a_{11}=(1-a_{00}-a_{33})/2$ and the chain rule of derivative. Owing
to $\langle n_{\bm{k}\uparrow}\rangle+\langle
n_{-\bm{k}\downarrow}\rangle=1/2-(a_{00}-a_{33})/2$, what concerns
our discussion is the value of $a_{00}-a_{33}$. Eq.~(\ref{free})
gives relation
\begin{equation}
1+(a_{00}-a_{33})^{2}-2(a_{00}+a_{33})=0.
\end{equation}
Substituting it in $\partial{F}/\partial{a_{00}}$, we have
\begin{equation}
\frac{\partial{F}}{\partial{a_{00}}}=-E_{\bm{k}}+Tk_{B}\ln
\frac{1+a_{00}-a_{33}}{1-(a_{00}-a_{33})}
\end{equation}
We see that Eq.~({\ref{first}}) is a transcendental equation of
$a_{00}-a_{33}$. To obtain an approximate solution, we consider the
first order expansion of $a_{00}-a_{33}$ with limit of strong
interaction, a large $\eta$, which leads to an approximate solution
\begin{equation}
a_{00}-a_{33}\simeq E_{\bm{k}}(\frac{\eta f\omega^{2}}{hf
E_{\bm{k}}^{2}+2\eta h\Delta^{2}}+2k_{B}T)^{-1}.
\end{equation}
Our approximation requires $a_{00}-a_{33}\ll 1$. From the solution,
this condition is ensured when $\Delta$ is a small number,
$\Delta\sim0$. When $\omega\sim0$, a small $E_{\bm{k}}\sim0$ makes
$a_{00}-a_{33}$ a small number. For $\omega\nsim0$, a large $\eta$
results in a small $a_{00}-a_{33}$.  It is necessary to note that
the solution not only applies to the gapped momentum, but also to
the states ungapped. This is because the non-unitary evolution
affects on all momentums as long as some momentums are gapped. For
the gapped momentums, $\Delta$ equals the gap constant, while
$\Delta=0$ for those ungapped momentums.

According to our assumption, the gapping occurs only in the subspace
where a gap reduces the energy. For an infinitesimal gap, the gapped
momentum should satisfy $\left.\frac{d(\delta
E_{\bm{k}})}{d(\Delta^{2})}\right|_{\Delta^{2}=0}<0$. With our
approximation, the derivative is given by
\begin{equation}
\left.\frac{d(\delta
E_{\bm{k}})}{d(\Delta^{2})}\right|_{\Delta^{2}=0}\simeq
\frac{1}{2\omega}-\frac{2h}{f}.
\end{equation}
For momentums near the Fermi surface, we have $\omega\sim0$, so the
gap for them can not be opened. With proper parameters $f$ and $h$,
momentums having a distance to the Fermi surface can be gapped.
Because the ungapped states can be excited without overcoming an
energy gap, the overall state is not in the superconducting phase.
Thus the pseudogap phenomenon emerges.

\section{Conclusion}
We studied the non-unitary evolution caused by the fluctuation
around the mean field average. We showed that such an evolution can
break the superconducting gap for momentums near the Fermi surface
and lead to the pseudogap phenomenon. Because the equilibrium state
is a non-canonical one, dynamics of the thermal relaxation also
influence its form. In the non-canonical equilibrium states, the
particle number does not obey the Fermi-Dirac distribution even in
non-interaction Fermion systems.

\acknowledgments The author is grateful to X. Sun, C.Q. Wu and Y.L.
Ma for useful discussions. The work was partially supported by the
Doctoral Foundation Program of the Chinese Ministry of Education.

\section{Appendix: Goal of thermalization}
Suppose an equilibrium ensemble of an $N$ dimensional system is
constituted by $|n\rangle=c_{nm}|E_{m}\rangle$, where
$|E_{m}\rangle\ (m=1,2,\cdots,N)$  is energy eigenstates. According
to the principle of minimum free energy, the equilibrium state is
given by
\begin{equation}
\rho=\frac{\textrm{exp}(-\beta\sum_{m}|c_{nm}|^{2}E_{m})|n\rangle\langle
n|}{\textrm{Tr}[\textrm{exp}(-\beta\sum_{m}|c_{nm}|^{2}E_{m})|n\rangle\langle
n|]}
\end{equation}
Then, taking the partial derivatives with respect to $c_{nm}$, one
can find the canonical equilibrium state ($a_{nm}=\delta_{nm}$) is a
local minimum of the free energy, while other states are not in
general. Of course, there may be other local minimums for delicate
assignment of the parameters, and a rigid argument should show the
canonical ensemble is the global minimum. In view of the common
validity of the canonical ensemble, however, we assume it is the
goal of thermal relaxation.

\end{document}